\documentclass[12pt]{article}
\usepackage{epsfig}
\usepackage{latexsym}
\usepackage{color}
\newcommand{\mysquare}[0]{\raise-.2ex\hbox{{\Large$\Box$}}}
\def\lsim{\mathrel{\rlap {\raise.5ex\hbox{$ < $}}
{\lower.5ex\hbox{$\sim$}}}}
\def\gsim{\mathrel{\rlap {\raise.5ex\hbox{$ > $}}
{\lower.5ex\hbox{$\sim$}}}} \topmargin -1.5cm \textheight=22.5cm
\catcode`\@=11
\newcount\hour
\newcount\minute
\newtoks\amorpm
\hour=\time\divide\hour by60
\minute=\time{\multiply\hour by60 \global\advance\minute by-\hour}
\edef\standardtime{{\ifnum\hour<12 \global\amorpm={am}%
        \else\global\amorpm={pm}\advance\hour by-12 \fi
        \ifnum\hour=0 \hour=12 \fi
        \number\hour:\ifnum\minute<10 0\fi\number\minute\the\amorpm}}
\edef\militarytime{\number\hour:\ifnum\minute<10 0\fi\number\minute}
\def\draftlabel#1{{\@bsphack\if@filesw {\let\thepage\relax
   \xdef\@gtempa{\write\@auxout{\string
      \newlabel{#1}{{\@currentlabel}{\thepage}}}}}\@gtempa
   \if@nobreak \ifvmode\nobreak\fi\fi\fi\@esphack}
        \gdef\@eqnlabel{#1}}
\def\@eqnlabel{}
\def\@vacuum{}
\def\draftmarginnote#1{\marginpar{\raggedright\scriptsize\tt#1}}
\def\draft{\oddsidemargin -.2truein
        \def\@oddfoot{\sl preliminary draft \hfil
        \rm\thepage\hfil\sl\today\quad\militarytime}
        \let\@evenfoot\@oddfoot \overfullrule 3pt
        \let\label=\draftlabel
        \let\marginnote=\draftmarginnote
   \def\@eqnnum{(\theequation)\rlap{\kern\marginparsep\tt\@eqnlabel}%
\global\let\@eqnlabel\@vacuum}  }

\relax

\newcommand{\ba}[0]{\begin{eqnarray}}
\newcommand{\ea}[0]{\end{eqnarray}}
\def\bs{\begin{subequations}}
\def\es{\end{subequations}}

\def\thebibliography#1{%
\vskip 0.5cm \centerline{\bf References}
\list{%
[\arabic{enumi}]}{\settowidth\labelwidth{[#1]}
\leftmargin\labelwidth
\advance\leftmargin\labelsep
\usecounter{enumi}}
\def\newblock{\hskip .11em plus .33em minus .07em}
\sloppy\clubpenalty4000\widowpenalty4000
\sfcode`\.=1000\relax}

\renewcommand{\theequation}{\arabic{section}.\arabic{equation}}

\renewcommand{\section}{\setcounter{equation}{0}\@startsection%
{section}{1}{0mm}{-\baselineskip}{0.5\baselineskip}%
{\normalfont\normalsize\bfseries}}

\renewcommand{\subsection}{\@startsection%
{subsection}{2}{0mm}{-\baselineskip}{0.5\baselineskip}%
{\normalfont\normalsize\bfseries}}

\renewcommand{\subsubsection}{\@startsection%
{subsubsection}{3}{0mm}{-\baselineskip}{0.5\baselineskip}%
{\normalfont\normalsize\slshape}}
\def\crbig{\\\noalign{\vspace{3mm}}}

\usepackage{amscd} \usepackage{amsmath} \usepackage{amsfonts}
\def\crbig{\\\noalign{\vspace{3mm}}}

\def\Re{\,{\rm Re}\, }

\def\s{\sigma}

\def\thefootnote{\fnsymbol{footnote}}
\def\es{\end{subequations}}

\def\ec{\hat E^{c}_{2}}

\newcommand{\uarrw}[0]{\mathrel{
{\raise.5ex\vbox{\hrule width 1cm}\hskip-6pt\rightarrow}}}

\def\bea{\begin{array}}
\def\bem{\begin{displaymath}}
\def\beq{\begin{equation}}

\def\eea{\end{array}}
\def\eem{\end{displaymath}}
\def\eeq{\end{equation}}

\def\ov{\overline}

\def\Re{\mathop{\rm Re}}

\def\s2w{\sin^2 \theta_W}

\def\crbig{\\\noalign{\vspace {3mm}}}

\def\be{\begin{equation}}
\def\ee{\end{equation}}
\def\bc{\begin{center}}
\def\ec{\end{center}}
\def\bea{\begin{eqnarray}}
\def\eea{\end{eqnarray}}

\begin{document}
\renewcommand{\theequation}{\arabic{section}.\arabic{equation}}
\begin{titlepage}
\begin{flushright}
NEIP--06--03 \\
LPTENS--06/03\\
CPTH--RR011.0206\\
hep-th/0602111 \\
\end{flushright}
\begin{centering}
\vspace{15pt}
\boldmath
{\bf
FLUXES, GAUGINGS AND GAUGINO CONDENSATES}$^\ast$\\
\unboldmath
\vspace{20pt}
{J.-P.~DERENDINGER$^1$,
C.~KOUNNAS$^{2}$,
\vskip .1cm
and
\vskip .1cm
P.M.~PETROPOULOS$^3$} \vskip .3cm {\small
$^1$ Physics Institute, Neuch\^atel University, \\
Breguet 1, CH--2000 Neuch\^atel, Switzerland \vskip .1cm $^2$
Laboratoire de Physique Th\'eorique,
Ecole Normale Sup\'erieure$^\dagger$, \\
24 rue Lhomond, F--75231 Paris Cedex 05, France \vskip .1cm $^3$
Centre de Physique Th\'eorique, Ecole Polytechnique$^\diamond$,
\\
F--91128 Palaiseau, France }

\vspace{20pt}

{\bf Abstract}\\
\end{centering}

\noindent Based on the correspondence between the $N=1$
superstring compactifications with fluxes and the $N=4$ gauged
supergravities, we study effective $N=1$ four-dimensional
supergravity potentials arising from fluxes and gaugino
condensates in the framework of orbifold limits of (generalized)
Calabi--Yau  compactifications. We give examples in heterotic and
type II orientifolds in which combined fluxes and condensates lead
to vacua with small supersymmetry breaking scale. We clarify the
respective roles of fluxes and condensates in supersymmetry
breaking, and analyze the scaling properties of the gravitino
mass.

\vspace{15pt} \noindent \small{\textsl{To appear in the
Proceedings of the {\it ``Corfu School and Workshops on
High-Energy Physics''}, Corfu, Greece, September 4 -- 26 2005.}}
\vfill \hrule width 6.7cm \vskip.1mm{\small \small \small
\noindent $^\ast$\ Research partially supported by the EU under
the contracts MEXT-CT-2003-509661,
MRTN-CT-2004-005104 and MRTN-CT-2004-503369.\\
$^\dagger$\ Unit{\'e} mixte  du CNRS et de l'Ecole
Normale Sup{\'e}rieure, UMR 8549.\\
$^\diamond$\  Unit{\'e} mixte  du CNRS et de  l'Ecole
Polytechnique, UMR 7644.}
\end{titlepage}
\newpage
\setcounter{footnote}{0}
\renewcommand{\thefootnote}{\arabic{footnote}}

\setlength{\baselineskip}{.7cm}
\setlength{\parskip}{.2cm}

\setcounter{section}{0}
\section{Introduction}

Gaugino condensation provides a way to induce nonperturbative
supersymmetry breaking in $N=1$ four-dimensional vacua. This
phenomenon occurs in the infrared regime of strongly coupled gauge
sectors \cite{gaugino1,gaugino2} and affects the superpotential of
the effective supergravity. These nonperturbative contributions
coexist, in the effective superpotential, with the perturbative
and nonperturbative moduli-dependent terms produced in heterotic
\cite{gaugino2, hetflux}, type IIA \cite{IIAflux} and type IIB
\cite{IIBflux} compactifications with fluxes.

The inclusion of the nonperturbative corrections in the
flux-induced superpotential has been proposed by several authors
\cite{condflux}. However, the conclusions have been either
controversial or incomplete, mainly due to the pathological
behaviour of the vacuum due \textit{e.g.} to a runaway behaviour
of the moduli involved in the condensate or to a fine-tuning
problem associated with the quantization of the flux coefficients.
Destabilization of the no-scale structure with undesired
transitions to anti-de Sitter vacua are also usual caveats.

In order to overcome the above difficulties, we must treat
simultaneously the flux and the condensate contributions in a
formalism which allows to capture unambiguously the corresponding
effects in the superpotential. Fluxes are data given at the
superstring level. Their translation into an $N=1$ effective
superpotential is ensured by the gauging procedure of some algebra
allowed by the massless content of the theory \cite{DKPZ, ACFI}.
Having a generic and unambiguous structure of the effective
superpotential in the presence of fluxes and gaugino condensates,
one can show that the usual pathologies of the vacuum are indeed
avoided in heterotic, IIA and IIB $N=1$ effective supergravities
\cite{DKP}. This is possible provided one realizes that the issues
of moduli stabilization, supersymmetry breaking, gaugino
condensation and positivity of the potential, although related,
must be treated \emph{separately}. A straightforward corollary of
this observation is that the nonperturbative contributions
\emph{are not always} the source for supersymmetry breaking, which
affects the scaling properties of the gravitino mass.

The present contribution is a summary of
the above results.

\section{Fluxes, gaugings and the description of condensates}\label{gen}

Even with broken supersymmetry, the underlying ten-dimensional
theory encodes the constraints of $N\geq 4$ supersymmetry, which
can then be used to derive information on the structure of the
effective $N = 1$ supergravity. In $N\geq 4$ supergravities, the
only available tool for generating a potential is to turn abelian
gauge symmetries, naturally associated with vector fields, into
non-abelian ones. This procedure of \emph{gauging} introduces in
the theory a gauge algebra $G$ acting on the vector fields in the
gravitational and/or vector supermultiplets \cite{drscwe}. The
important fact is that from the point of view of the ``daughter"
$N = 1$ supergravity obtained after orbifold and/or orientifold
projection, the gauging modifications only affect the
superpotential $W$, whereas the K\"ahler potential $K$ and hence
the kinetic terms remain the same as in the ungauged theory.

We will focus here on a $Z^2\times Z^2$
orbifold projection in heterotic
(plus orientifold $\Omega$ projection in type II),
which leads to the following K\"ahler manifold:
\begin{equation}
\mathcal{K} = \left( { SU(1,1) \over U (1)} \right)_{S}
 \times \prod_{A = 1}^{3}
\left( {SO(2,2 + n_A) \over SO (2) \times
SO(2+n_A)}\right)_{T_A,U_A,Z^I_A}.
\end{equation}
Each string compactification is characterized by its own
parameterization of the scalar manifold in terms of the seven
complex scalars  $S,T_A,U_A,$ $A=1,2,3$ and the matter scalar
fields $Z^I_A$. In the case of heterotic, the index $A$ labels the
three complex planes defined by the $Z_2 \times Z_2$ symmetry used
for the orbifold projection. For type II compactifications, this
holds up to field redefinitions, which mix all $S,T_A$ and $U_A$.
The structure of the scalar manifold remains however unaltered.

This choice of parameterization singles out the appropriately
redefined geometric moduli and the dilaton as $\Re T_A$, $\Re
U_A$, $\Re S$. Neglecting for simplicity the matter scalar fields
$Z^I_A$, the manifold of the geometric moduli is reduced to
\begin{equation}
\mathcal{K} = \left( { SU(1,1) \over U (1)} \right)_{S}
 \times \prod_{A = 1}^{3}
\left( {SU(1,1) \over U(1)}\right)_{T_A} \times\prod_{A = 1}^{3}
\left( {SU(1,1) \over U(1)}\right)_{U_A}.
\end{equation}
The K\"ahler potential associated to these field takes for all
cases the following form:
\begin{equation}
K = - \log \left(S+{\ov S}\right) -\sum_{A=1}^3 \log
\left(T_A+{\ov T}_A\right)\left(U_A+{\ov U}_A\right).
\end{equation}
Due to the $SU(1,1)^7$-structure of the manifold, the scalar
potential considerably simplifies and takes the following
suggestive form,
\begin{equation}
\label{pot3} {\rm e}^{-K} V = \sum_i \left| W - W_i(Z_i + \ov
Z_i)\right|^2 - 3 |W|^2,
\end{equation}
where $ \{ Z_i \} \equiv \{ S, T_A, U_A \}$ and $W_i =
\partial_{Z_i } W$.

The structure of the superpotential is also well understood. The
perturbative part of the latter is polynomial in the moduli with
coefficients related to the fluxes of the underlying string
theory. The simplicity of the approach based on gauging of the
underlying $N=4$ supersymmetry algebra allows to study
exhaustively various situations and establish a precise dictionary
among monomial coefficients in the superpotential and fluxes
(including also spin-connection geometric fluxes). The precise
analysis can be found in \cite{DKPZ} where we concentrated on
orientifolds of type IIA strings which offered the broadest
structure of allowed fluxes and had been explored to a lesser
extent \cite{IIAflux}.

The non-perturbative effects originating from gaugino condensation
provide modifications in the superpotential of the effective
supergravity theory. These are of the form
\begin{equation}
\label{wnp1} W_{\rm nonpert} = \mu^3 \, {\rm exp}\left( - {24\pi^2
Z \over b_0 } \right),
\end{equation}
where $b_0$ is a one-loop beta-function coefficient, $\mu$ a scale
at which the Wilson coupling $g^2(\mu)$ is defined and $Z$ a
modulus such that $\Re Z = g^{-2}(\mu)$. The expectation value of
the nonperturbative superpotential defines the
renormalization-group-invariant transmutation scale $\Lambda$ of
the confining gauge sector in which gauginos condense, $\langle
W_{\rm nonpert.} \rangle= \Lambda^3$. The nature of the modulus
$Z$ depends on the underlying string (or M-) theory
compactification: in the heterotic string, it is identified with
the dilaton $S$ field, whereas in type II orientifolds $Z$ is the
redefined $S$ field in IIA theories \cite{DKPZ} and a combination
of $T$ and $S$ in IIB (or F-theory) compactifications
\cite{gaugino2, DKPZ,het,LRS}. The resulting exponent is a number
of order ten or more and $n$-instanton corrections ($n>1$) are
exponentially suppressed.

Two remarks are in order here. First, many gaugino condensates
could form and the nonperturbative superpotential could include
several similar terms involving various moduli. Second, the scale
$\mu$ in (\ref{wnp1}) is in general a modulus-dependent quantity.

We will not expand any longer on the general structure of the
nonperturbative contributions to the superpotential. A
comprehensive analysis can be found in \cite{BDFS}. We will
instead focus on a simple situation where the superpotential
reads:
\begin{equation}
\label{ftW}
W = a + w(S),
\qquad\qquad
w(S)= \mu^3 {\rm e}^{-S}
\end{equation}
($S$ has been rescaled according to $24\pi^2 S/b_0 \to S$, which
leaves the corresponding kinetic terms unchanged and multiplies
the scalar potential by an overall factor). The scale $\mu$ and
the quantity $a$ are in general moduli-dependent; $a$ includes the
perturbative contributions induced by fluxes\footnote{It does not
depend on $S$ in heterotic compactifications.}.

\section{The fine-tuning and the runaway problems}\label{ftr}

Consider now the situation in heterotic where the geometrical
fluxes are absent and thus the superpotential is
$T_A$-independent. The scalar potential becomes:
\begin{equation}
\label{nspot} {\rm e}^{-K} V = \sum_{\{ Z_i \} \equiv \{ S, U_A
\}} \left|W - W_i(Z_i + \ov Z_i)\right|^2.
\end{equation}
This exhibits a \emph{no-scale} structure \cite{noscale}, with a
semi-positive-definite potential and flat directions $\{T_A\}$.
The $U_A$ moduli are generically fixed by their minimization
conditions and $a$ and $\mu^3$ in Eqs.~(\ref{ftW}) are effectively
constant. The remaining minimization condition for the $S$ field,
\begin{equation}
\label{minS} a + \left(S+\ov S + 1\right)w(S)=0,
\end{equation}
determines the value of $S$. Supersymmetry is broken in the
$T_A$-directions, in Minkowski space. Equation (\ref{minS}) shows
that an exponentially small value of $w(S)$ necessarily implies $
| a | \ll 1$, a stringent \emph{fine-tuning} condition. This is a
severe problem in situations (such as the $Z_3$ orbifold in
heterotic) where $a$ is directly given by constant perturbative
fluxes, hence it must either vanish or be of order one. A
vanishing $a$ is also problematic because it leads to a
\emph{runaway} potential, $V \propto |\mu|^6 \exp{-\left(S+\ov
S\right)}$. The attempts that have been proposed in the past for
improving this situation include \textit{e.g.} multiple gauge
group condensations (without fluxes). These do not help in
removing the fine-tuning problem with a non-zero $a$.

The fine-tuning and runway problems are facets of the vacuum
structure of the theory. As such, they can be understood
\emph{only} from a comprehensive analysis of the combined
perturbative and nonperturbative contributions, which equally
contribute in the structure of the vacuum. In other words, it
makes little sense to focus first on stabilizing the moduli from
the flux superpotential, and add the condensate contributions in a
second stage. Usually the first-stage flux superpotential turns
out to be ``too stable'' to lead to relevant phenomenology once
condensates are added.

One can illustrate the above in a type IIB model with
superpotential
\begin{eqnarray}\label{kacpol}
W&=& A\left[ 1 +U_1U_2+U_2U_3+U_3U_1 + \gamma  S(U_1+U_2+U_3 +
U_1U_2U_3)\right] \nonumber \\ &+&i B \left[ U_1+U_2+U_3 +
U_1U_2U_3 + \gamma S(1+ U_1U_2+U_2U_3+U_3U_1) \right],
\end{eqnarray}
Here $A, B$ and $\gamma A, \gamma B$ are respectively proportional
to R-R and NS-NS flux numbers. The no-scale structure with
semi-positive-definite scalar potential is due to the absence of
any $T_A$ dependence. The moduli $(U_A, \gamma S)$ are fixed to
unity and supersymmetry is broken in flat space with gravitino
mass $m_{3/2}^2 \propto (A^2 +B^2)/\prod_A(T_A + \ov T_A)$.
Nonperturbative contributions to the superpotential (\ref{kacpol})
may originate from $D_3$- or $D_7$-branes. In the latter case,
these contributions are of the form $\exp(-\alpha T)$ and their
presence spoils \emph{arbitrarily} the no-scale structure and
destabilizes the Minkowski vacuum: the moduli $T$ gets stabilized
but the potential becomes negative, $V = - 3m_{3/2}^2$, as
required by unbroken supersymmetry in anti-de Sitter space.

\section{Stationary points and Minkowski vacuum} \label{secVmin}

When supersymmetry breaks, the analysis of the non-positive scalar
potential as a function of seven complex fields is difficult. It
is somewhat simpler under the assumption of vanishing of the
potential at the minimum. In a general supergravity theory with
K\"ahler potential $K = - \sum_j\ln( Z_j + \ov Z_j)$,
supersymmetry is spontaneously broken if the equations
 \beq \label{Fjis} F_j  \equiv W - (Z_j + \ov Z_j )
 W_j =0
 \eeq
cannot be solved for all scalar fields $Z_j$ (and with $\Re
Z_j>0$). If supersymmetry breaks in Minkowski space, we have also
 \beq \label{minbr}
 \langle V \rangle =
 0 \, , \qquad\qquad \langle W \rangle \ne 0 \,.
 \eeq
A stationary point of the scalar potential is a solution of the
equation $\partial_j V =0$, for each scalar field $Z_j$. The
explicit analysis of these equations, under the assumptions
(\ref{minbr}), can be worked out explicitly. The scalar fields
split in two categories for which we use lower- ($a,b, \dots$) and
upper-case ($A,B, \dots$) indices respectively: either $\langle
W_a\rangle =0$ and $\langle F_a\rangle = \langle W \rangle \ne 0$,
or $\langle F_A\rangle =0$. Supersymmetry breaking is controlled
by the first category only. The Minkowski condition, $\langle
V\rangle =0$, implies then that this category contains precisely
\emph{three} fields: the contribution of each of these fields to
$\langle V \rangle$ cancels one unit of the negative term
$-3\langle W\ov W\rangle$. The seven minimization equations
finally read:
  \beq\label{Vminall}
  0 = \sum_{a=1}^3  W_{aj}\Re Z_a \quad
  \forall j \ (a\ \mathrm{or} \ A) ,
  \eeq
(the summation is restricted over moduli which break supersymmetry
i.e. with $\langle W_a\rangle =0$).

\section{ Supersymmetry breaking independent of the
gaugino condensation}\label{sbsg}

Nonperturbative phenomena do not necessarily break supersymmetry,
independently of their effect on the stabilization of the moduli
and on the positivity of the potential. The introduction of fluxes
can indeed stabilize some of the moduli in flat space without
inducing supersymmetry breaking. To be concrete, let us consider a
superpotential with ``supersymmetric mass terms" only:
 \beq
 \label{susy1}
 \begin{array}{rcl}
 W_{\rm susy} &=& A (U_1-U_2) (T_1 - T_2) +
 B (U_1 + U_2 - 2 U_3 ) ( T_1 + T_2 - 2 T_3)
 \crbig && + (T_1+T_2-2T_3) \, w(S) ,
 \end{array}
 \eeq
with $w(S)$ as given in Eq.~(\ref{ftW}). The perturbative part of
this superpotential is created by geometrical fluxes, either in
heterotic or in type IIA. It can be directly generated at the
string level using freely acting orbifold constructions. It
selects four (complex) directions in the seven-dimensional space
of the moduli fields and minimizing the potential tends to cancel
the fields in these four directions. Supersymmetry is not broken
and cancellation of auxiliary fields ($\langle F_A\rangle=0$)
fixes $U_A =U$ and $T_A =T$.

The condensate term $(T_1 + T_2 - 2 T_3) w(S)$ can be understood
from the general form of the $N=4$ superpotential (for details,
see \cite{DKP}). The presence of $w(S)$ leaves $U_1 = U_2 = U$ but
we now have $U_3=U  + w(S)/2B $. The above conclusions about
supersymmetry remain however unchanged: supersymmetry is unbroken
and the gravitino is massless in flat background.

The previous example can be modified by the addition of further
flux terms which break supersymmetry. Consider for example, in the
heterotic or type IIA,
 \beq \label{het1} W_{\rm total} = W_{\rm
 susy} + W_{\rm break} \eeq with $W_{\rm susy}$ as in
 Eq.~(\ref{susy1})
 and
 \be
 W_{\rm break}= R\,(T_1 U_1 + T_2 U_2).
 \ee
The term $W_{\rm break}$ breaks supersymmetry \emph{even in the
absence} of $w(S)$. The scalar potential has a minimum with real
$T_A$, $U_A$ and $T_A = T$, $U_1 = U_2 = U$, $U_3 = U + w(S)/2B$.
The potential vanishes along the flat directions $S$, $T$ and $U$.
The goldstino field is a combination of the fermionic partners of
$S, T_3$ and $U_3$. There is an ``effective'' no-scale structure:
since $W_i$ vanishes in these three directions, their
corresponding contributions to the potential cancel the
gravitational contribution $-3m_{3/2}^2$. Thus supersymmetry is
broken in flat space--time with
$$
m_{3/2}^2 = {|R|^2\over 32 \, ST_3U_3}
$$
and the presence of the nonperturbative term $w(S)$ only acts as a
\emph{small perturbation} on supersymmetry breaking induced by the
modulus-dependent contribution $W_{\rm break}$. It however
explicitly appears in mass terms.

Many other examples can be worked out in the same line of thought
in heterotic, type IIA and type IIB, where the role of the gaugino
condensate is not crucial for the supersymmetry breaking. An
explicit breaking term, generated by a specific combination of
fluxes, has to be superimposed to the mass terms, in order for the
supersymmetry to be broken, independently of the presence of the
condensate. This situation is not generic, however, and examples
exist where \emph{supersymmetry breaking is triggered by the
gaugino condensate}.

\section{Gaugino-induced supersymmetry breaking}\label{sbg}

We will now analyze situations where the gaugino condensate breaks
supersymmetry with $m_{3/2}$ being related to the gaugino scale
$w(S)$. In these cases the form of the superpotential will be
again
 \beq
 W= W_{\rm susy} + \mu^3(Z_i) \, {\rm e}^{-S},
 \eeq
but $\mu^3(Z_i)$ will no longer vanish at the minimum.

\subsection*{Type II}\label{sbag}

Consider the following type IIA superpotential
\begin{equation}
\label{WIIAsb} W= \left(T_1 - T_2\right)\left(- U_1+ U_2  - T_3  +
2S \right) + \left(U_1 T_3 -L\right)w(S),
\end{equation}
which is generated by geometric and $F_2$ fluxes. It falls in the
class mentioned in Sec. \ref{secVmin}, where there is a partition
between directions which break supersymmetry (here $T_1$, $T_2$
and $U_3$) and directions which preserve supersymmetry ($T_3, U_1,
U_2$ and $S$).

The  requirement $\langle W_{T_1} \rangle = \langle W _{T_2}
\rangle= 0$ ensures $\langle V \rangle =0$ since $W$ is
independent of $U_3$, and the resulting supersymmetry-breaking
condition reads
\begin{equation}
\label{sb1}   - U_1 + U_2- T_3 +2 S =0.
\end{equation}
The vanishing of the $F$-auxiliary fields in the directions
$T_3,U_1,U_2$ and $S$ leads to the following equations:
\begin{eqnarray}
\xi\left(\ov U_1 + U_2 - T_3 + 2  S\right) - \left( \ov
 U_1 T_3 + L\right)w(S)&=&0, \label{co1IIAsb1}\\
\xi\left(- U_1  -\ov U_2 - T_3 + 2  S\right) + \left(
 U_1 T_3 - L\right)w(S)
&=&0,\label{co2IIAsb1} \\
\xi\left(- U_1  + U_2 + \ov T_3 + 2  S\right) - \left(
 U_1 \ov T_3 + L\right)w(S)
&=&0, \label{co3IIAsb1}\\
\xi\left( - U_1  + U_2 - T_3 - 2  \ov S\right) + \left(
 U_1 T_3 - L\right)\left(
 1 + S + \ov S\right)w(S)
&=&0,\label{co4IIAsb1}
\end{eqnarray}
where we have introduced
 \begin{equation}\label{xi}
 \xi \equiv T_1 - T_2.
 \end{equation}
The minimization condition (\ref{Vminall}) reads here
\begin{equation}\label{minxi}
\Re  \xi =0.
\end{equation}

Equations (\ref{sb1})--(\ref{minxi}) must be solved for $\xi, T_3,
U_1, U_2$ and $S$. Combining Eqs.~(\ref{co1IIAsb1}) and
(\ref{co3IIAsb1}), one concludes that $T_3 = U_1$. A similar
combination of Eqs.~(\ref{co1IIAsb1}) and (\ref{co2IIAsb1}) shows
that these moduli must be chosen real: $T_3 = U_1 = t$. The
requirement (\ref{minxi}) can be fulfilled by adjusting
appropriately the imaginary part of the $S$ field: $S = s -
\frac{i}{2}(\pi - 6 \varphi_\mu)$ (we have introduced $\mu =
|\mu|\exp i \varphi_\mu $). This implies through Eq.~(\ref{sb1})
that $U_2 = u + i (\pi - 6 \varphi_\mu)$. The final equations for
$t,u$ and $s$ are (\ref{sb1}), a combination of (\ref{co1IIAsb1})
and (\ref{co2IIAsb1}) as well as a combination of (\ref{sb1}),
(\ref{co1IIAsb1}) and (\ref{co4IIAsb1}):
 \begin{eqnarray}
 u+2(s-t)
 &=&0, \label{1IIAsb1}\\
  {t}\left(t^2-L\right)- {u}\left(t^2+L\right)
 &=&0, \label{2IIAsb1}\\
 t^5+ 2 L t^3- 4 L t^2- 3 L^2 t-4 L^2 &=&0. \label{3IIAsb1}
 \end{eqnarray}

We will not reproduce the full analysis of these equations here,
but instead state the results (details are available in
\cite{DKP}). Assuming the flux number $L$ large, the leading and
sub-leading behaviour for $t$, $u$ and $s$ is
\begin{eqnarray}
t&=& \sqrt{L} + 1 + \mathcal{O}\left(\frac{1}{\sqrt{L}}\right), \label{leadtLsb1}\\
u&=& 1 + \mathcal{O}\left(\frac{1}{L}\right),\label{leaduLsb1} \\
s&=& \sqrt{L} + \frac{1}{2} +
\mathcal{O}\left(\frac{1}{\sqrt{L}}\right).\label{leadsLsb1}
\end{eqnarray}
Next one can compute
 \beq {\rm Im} \, \xi \approx \sqrt{L}\, |\mu|^3 \, {\rm
 e}^{-\sqrt{L}}, \eeq
which measures the supersymmetry breaking. The gravitino mass
scales as
 \beq {\rm
 e}^{-K/2} \, m_{3/2} \approx 2i \sqrt{L}\, |\mu|^3 \, {\rm
 e}^{-\sqrt{L}}. \eeq
The last two equations show that the gaugino condensate is
entirely responsible for the breaking of supersymmetry. Notice
also that as advertised previously, the fluxes generating the
superpotential (\ref{WIIAsb}) are not fine-tuned, and solutions
for the moduli exist generically.

\subsection*{Heterotic}\label{shet}

Because of the absence of perturbative $S$-contributions in the
heterotic superpotential, heterotic and type II are drastically
different when the breaking of supersymmetry is induced by a
gaugino condensate. Let us concentrate on a superpotential of the
type
\begin{equation}\label{Whet}
 W= {\hat A} U_1 + {\hat B} U_2 +{\hat C} U_3 + {\hat D} U_4,
\end{equation}
where $U_4 = U_1 U_1 U_3$. This superpotential is odd in the
$U_i$'s and captures most of the heterotic compactifications
considered here, with a gaugino condensate. We have introduced the
following functions of $T_1,T_2$ and $S$:
\begin{eqnarray}
 {\hat A}&=& \Big[\alpha + \alpha' w(S)\Big]\xi + Aw(S), \label{WhetA}\\
 {\hat B}&=& \Big[\beta + \beta' w(S)\Big]\xi + Bw(S), \label{WhetB}\\
 {\hat C}&=& \Big[\gamma + \gamma' w(S)\Big]\xi + Cw(S), \label{WhetC}\\
 {\hat D}&=& \Big[\delta + \delta' w(S)\Big]\xi + Dw(S),\label{WhetD}
\end{eqnarray}
where $\xi = T_1-T_2$ as defined in (\ref{xi}) and $w(S)$ in
(\ref{ftW}).

The minimization condition (\ref{Vminall}) reads $\Re \xi =0$, as
in the above type IIA example. We will therefore choose $S= s
-i{(\pi - 6 \varphi_\mu)/2}$ and $U_i= u_i$ real. Everything is
consistent provided $\alpha,\beta,\gamma,\delta$ and $A,B,C,D$ are
real and $\alpha',\beta',\gamma',\delta'$ are imaginary.

The no-scale requirement $\langle V \rangle =0$ is fulfilled
provided $\langle W_{T_1} \rangle = \langle W _{T_2} \rangle= 0$
($W$ is independent of $T_3$). The corresponding condition reads:
\begin{equation}\label{sbhet}
  (\alpha + \alpha' w)u_1 +(\beta + \beta' w)u_2 + (\gamma +
\gamma'w)u_3 +(\delta + \delta' w)u_4=0.
\end{equation}
The vanishing of the $U_A$--auxiliary fields leads to
\begin{eqnarray}
 -{\hat A} u_1 + {\hat B} u_2 +{\hat C} u_3 -{\hat D} u_4&=&0 , \label{hetu1}\\
  {\hat A} u_1 - {\hat B} u_2 +{\hat C} u_3 - {\hat D} u_4&=&0 , \label{hetu2}\\
  {\hat A} u_1 + {\hat B} u_2 -{\hat C} u_3 - {\hat D} u_4&=&0  \label{hetu3}
\end{eqnarray}
and the equation for the $S$-auxiliary field (after some
simplification involving Eqs. (\ref{sbhet})--(\ref{hetu3})) reads:
\begin{equation}\label{hetssol}
  {2 \over s}=-4-\left( {\alpha '\over \hat A} + {\beta '\over \hat
B}+{\gamma '\over \hat C}+ {\delta '\over \hat D}\right)\xi w.
\end{equation}

The equations at hand can be solved. We will exhibit a solution in
the plane-symmetric situation, where
\begin{equation}
\alpha = \beta = \gamma\ , \ \  \alpha' = \beta' = \gamma'\ , \ \
A = B = C,
\end{equation}
which imply that $ {\hat A}={\hat B}={\hat C} $ and consequently
\begin{equation}\label{lamups}
u\equiv  u_1 = u_2 = u_3 = \sqrt{\hat A\over \hat D} \quad
\mathrm{and} \quad u_4=u^3.
\end{equation}
The final set of equations for $\xi$, $u$ and $s$ is therefore
(\ref{sbhet}), (\ref{hetssol}) and (\ref{lamups}). Eliminating $u$
from (\ref{sbhet}) (by using (\ref{lamups})) leads to
\begin{equation}\label{xieq}
4\xi=-{3D\alpha+A\delta+(3D\alpha'+A\delta')w \over
(\alpha+\alpha'w)(\delta+\delta'w)}\, w.
\end{equation}
The latter can be further used in Eq. (\ref{hetssol}) (together
with (\ref{WhetA})) to obtain the central equation for the
determination of $s$:
\begin{equation}\label{seq}
{2\over s}=-4-{(\alpha'\delta-\delta'\alpha)w\over
(\alpha+\alpha'w)(\delta+\delta'w)}{3D\alpha+A\delta + (3\alpha' D
+A\delta')w \over
 D\alpha-A\delta+(D\alpha'-A\delta')w
}.
\end{equation}

For further simplification we specialize to
\begin{equation}\label{spe}
\alpha'=i\alpha\ , \ \ \delta'=-i\delta.
\end{equation}
Our aim is to show that Eq.~(\ref{seq}) indeed admits physically
acceptable solutions for $s$, provided that the fluxes
$\alpha,\delta,A$ and $D$ are large, while their ratios (such as
$D\alpha/A\delta$) are of order unity. Under these assumptions, we
can perform an expansion in powers of $w$ for all quantities. We
find the following dominant contributions (Eqs.~(\ref{lamups}),
(\ref{xieq}) and (\ref{seq})):
\begin{eqnarray}
u&\approx & \sqrt{-3\alpha \over\delta}, \label{uapproxspe}\\
\xi&\approx & -{D\over \delta}~w, \label{xiapprox}\\
{A\delta-D\alpha\over D\alpha }& \approx & 2i{4s+1\over
2s+1}w(s).\label{sapproxspes}
\end{eqnarray}
The latter equation is compatible with large values of $s$. In
that regime it further simplifies:
\begin{equation}\label{sapproxspess}
s\approx\log\left({{4|\mu|^3 D\alpha\over D\alpha-A\delta
}}\right)-{1\over 4\log\left({{4|\mu|^3 D\alpha\over
D\alpha-A\delta }}\right)}.
\end{equation}

Using finally Eqs.~(\ref{Whet}), (\ref{WhetA}) and (\ref{lamups}),
in the special case captured by (\ref{spe}) and within the above
approximations, the gravitino mass reads:
\begin{equation}
{\rm e }^{-K/2}m_{3/2} \approx i4D\left(-
\frac{3\alpha}{\delta}\right) ^{3/2}\frac{s}{2s+1} w^2,
\end{equation}
with $s$ given in Eq.~(\ref{sapproxspess}). The gravitino mass
scales as $w^2$ instead of $w$ like in previous type II example.
This is due to the absence of flux-induced $S$-term in the
superpotential, which is a generic feature in heterotic
compactifications

\section{Conclusions}

The important outcome of the above analysis is that the
pathological behaviour of the vacuum in the presence of fluxes
with nonperturbative corrections \emph{is not} a generic property
of the $N=1$ effective supergravity, with or without spontaneously
broken supersymmetry: the usual caveats quoted previously can be
avoided with \emph{a suitable combination of fluxes and
nonperturbative contributions}. When appropriate, such a
combination is a valuable tool for circumventing the runaway
behaviour of moduli or the fine-tuning problem. Hence,
understanding these mechanisms can shed light on the nature of the
vacuum and the stabilization of moduli.

Our analysis makes also clear the importance of investigating  the
full superpotential, including both flux-induced perturbative and
nonperturbative contributions. Although nonperturbative
corrections do not necessarily trigger supersymmetry breaking,
they can alter various terms and must therefore be taken into
account at a very early stage: they can drastically change the
picture of moduli stabilization drawn by fluxes only.

On the practical side, the situations that we have investigated
fall in at least \emph{three} classes, according to the scaling
behaviour of the gravitino mass:
\begin{enumerate}
\item Situations where the nonperturbative corrections do not
  trigger the supersymmetry breaking -- they modify the mass terms though
  -- in which the gravitino mass scales like:
  $$m_{3/2} = c\,/ \sqrt{V} , $$
  where $V = \exp K$ is the volume of the moduli space and $c$
  is related to the flux numbers. In these cases, any mass hierarchy
  strongly relies on the volume $V$ of the moduli space. This
  behaviour can occur in all types of string compactifications,
  type IIA,B and heterotic.
\item  Situations where the nonperturbative contributions to the
  superpotential induce the supersymmetry breaking, where the gravitino mass
  is controlled by the nonperturbative superpotential $w(S)$, like the type IIA
  example presented in Sec. \ref{sbg},
  \begin{equation}\label{IIAhier}
  m_{3/2} = c\,w(S) \,/ \sqrt{V},
  \end{equation}
  as commonly expected from nonperturbative supersymmetry breaking.
  In these cases, the mass hierarchy can be created irrespectively of
  the size of the volume of the moduli space.
\item
The \emph{third} scaling behaviour is more unusual but still
generic in heterotic. It appears whenever the gaugino condensate
is the only source of $S$-dependence. The gravitino mass scales
now as
  \beq
  m_{3/2} = c\,w(S)^2\, / \sqrt{V}.
  \eeq
Hence, the supersymmetry breaking creates a mass hierarchy
stronger than in the two previous cases. The heterotic
realizations under consideration are actually quite generic
despite the fact that the ratios between flux coefficients are
required to be of order one, while the coefficients themselves are
large. However, it is clear that heterotic models deserve a more
systematic investigation, where gauge condensates of the type
$\left\langle \frac{1}{g^2}F \right\rangle$ are taken into
account, together with the gaugino ones. This should shed light on
the validity of various gaugino-condensate-induced supersymmetry
breaking scenarios advertised in the literature but not captured
by the analysis presented here.
\end{enumerate}

It should finally be stressed that the analysis of the $N=1$,
low-energy soft-supersymmetry-breaking terms strongly depends on
the class of model under consideration since their pattern is
mostly controlled by the radiative corrections induced by the
supersymmetry-breaking sector.

\section*{Acknowledgements}

We would like to thank the organizers of the {\it ``Corfu School
and Workshops on High-Energy Physics''} held in Corfu, Greece,
from September 4 to 26 2005, where these results were presented.
Based on various works supported in part by the EU under the
contracts MEXT-CT-2003-509661, MRTN-CT-2004-005104,
MRTN-CT-2004-503369, by the Swiss National Science Foundation and
by the Agence Nationale pour la Recherche, France.


%

\begin{thebibliography}{99}

\bibitem{gaugino1}
  S.~Ferrara, L.~Girardello and H.~P.~Nilles,
  Phys.\ Lett.\ B {\bf 125} (1983) 457.

\bibitem{gaugino2}
  J.-P.~Derendinger, L.~E.~Ib\'a\~nez and H.~P.~Nilles,
  Phys.\ Lett.\ B {\bf 155} (1985) 65;\\
  M.~Dine, R.~Rohm, N.~Seiberg and E.~Witten,
  Phys.\ Lett.\ B {\bf 156} (1985) 55.

\bibitem{hetflux}
A.~Strominger, Nucl.\ Phys.\ B {\bf 274} (1986) 253;\\
R.~Rohm and E.~Witten, Annals Phys.\  {\bf 170} (1986) 454; \\
N.~Kaloper and R.~C.~Myers, JHEP {\bf 9905} (1999) 010
[arXiv:hep-th/9901045].

\bibitem{IIAflux}
J.~Polchinski and A.~Strominger, Phys.\ Lett.\ B {\bf 388} (1996)
736 [arXiv:hep-th/9510227];\\
I.~Antoniadis, E.~Gava, K.~S.~Narain and T.~R.~Taylor, Nucl.\
Phys.\
B {\bf 511} (1998) 611 [arXiv:hep-th/9708075];\\
S.~Gukov, C.~Vafa and E.~Witten, Nucl.\ Phys.\ B {\bf 584} (2002)
69 [Erratum-ibid.\ B {\bf 608} (2001) 477] [arXiv:hep-th/9906070];\\
S.~Gukov, Nucl.\ Phys.\ B {\bf 574} (2000) 169
[arXiv:hep-th/9911011].

\bibitem{IIBflux}
J.~Michelson, Nucl.\ Phys.\ B {\bf 495} (1997) 127
[arXiv:hep-th/9610151];\\
K.~Dasgupta, G.~Rajesh and S.~Sethi, JHEP {\bf 9908} (1999) 023
[arXiv:hep-th/9908088];\\
T.~R.~Taylor and C.~Vafa, Phys.\ Lett.\ B {\bf 474} (2000) 130
[arXiv:hep-th/9912152];\\
P.~Mayr, Nucl.\ Phys.\ B {\bf 593} (2001) 99
[arXiv:hep-th/0003198];\\
G.~Curio, A.~Klemm, D.~L\"ust and S.~Theisen, Nucl.\ Phys.\ B {\bf
609} (2001) 3 [arXiv:hep-th/0012213];\\
S.~B.~Giddings, S.~Kachru and J.~Polchinski, Phys.\ Rev.\ D {\bf
66} (2002) 106006 [arXiv:hep-th/0105097];\\
S.~Kachru, M.~B.~Schulz and S.~Trivedi, JHEP {\bf 0310} (2003) 007
[arXiv:hep-th/0201028];\\
A.~R.~Frey and J.~Polchinski, Phys.\ Rev.\ D {\bf 65} (2002)
126009
[arXiv:hep-th/0201029]; \\
S.~Ferrara and M.~Porrati,
Phys.\ Lett.\ B {\bf 545} (2002) 411
[arXiv:hep-th/0207135]; \\
C.~Angelantonj, S.~Ferrara and M.~Trigiante, JHEP {\bf 0310}
(2003) 015 [arXiv:hep-th/0306185] and
Phys.\ Lett.\ B {\bf 582} (2004) 263 [arXiv:hep-th/0310136]; \\
 M.~Grana, T.~W.~Grimm, H.~Jockers and J.~Louis,
  Nucl.\ Phys.\ B {\bf 690}, 21 (2004) [arXiv:hep-th/0312232]; \\
  T.~W.~Grimm and J.~Louis,
  Nucl.\ Phys.\ B {\bf 699}, 387 (2004)
  [arXiv:hep-th/0403067]; \\
 P.~G.~Camara, L.~E.~Ib\'a\~nez and A.~M.~Uranga,
  Nucl.\ Phys.\ B {\bf 708} (2005) 268
  [arXiv:hep-th/0408036];\\
  H.~Jockers and J.~Louis,
  Nucl.\ Phys.\ B {\bf 705}, 167 (2005)
  [arXiv:hep-th/0409098].

\bibitem{condflux}
 G.L. Cardoso, G. Curio, G. Dall'Agata, D. L\"ust,
  Fortsch. Phys. \textbf{52} (2004) 483
  [arXiv:hep-th/0310021];\\
  G.L. Cardoso, G. Curio, G. Dall'Agata, D. L\"ust,
  JHEP \textbf{0409} (2004) 059
  [arXiv:hep-th/0406118];\\
  L.~Gorlich, S.~Kachru, P.~K.~Tripathy and S.~P.~Trivedi,
  JHEP {\bf 0412} (2004) 074
  [arXiv:hep-th/0407130];\\
  K.~Choi, A.~Falkowski, H.~P.~Nilles, M.~Olechowski and S.~Pokorski,
  JHEP {\bf 0411} (2004) 076
  [arXiv:hep-th/0411066];\\
 P.~Manousselis, N.~Prezas and G.~Zoupanos,
  arXiv:hep-th/0511122.

\bibitem{DKPZ}
  J.-P.~Derendinger, C.~Kounnas, P.~M.~Petropoulos and F.~Zwirner,
  Nucl.\ Phys.\ B {\bf 715} (2005) 211
  [arXiv:hep-th/0411276] and
  Fortsch.\ Phys.\  {\bf 53} (2005) 926
  [arXiv:hep-th/0503229].


\bibitem{ACFI}
G.~Aldazabal, P.~G.~Camara, A.~Font and L.~E.~Ibanez,
  arXiv:hep-th/0602089.

\bibitem{DKP}
  J.-P.~Derendinger, C.~Kounnas and P.~M.~Petropoulos,
  arXiv:hep-th/0601005.

\bibitem{drscwe}
  M.~de Roo and P.~Wagemans, Nucl.\ Phys.\ B {\bf 262} (1985) 644
  and
  Phys.\ Lett.\ B {\bf 177} (1986) 352;\\
 J.~Schon and M.~Weidner,
  arXiv:hep-th/0602024.

\bibitem{het}
  E.~Witten,
  Phys.\ Lett.\ B {\bf 155} (1985) 151;\\
  J.-P.~Derendinger, L.~E.~Ib\'a\~nez and H.~P.~Nilles,
  Nucl.\ Phys.\ B {\bf 267} (1986) 365.

\bibitem{LRS}
  D.~L\"ust, S.~Reffert and S.~Stieberger,
  Nucl.\ Phys.\ B {\bf 706} (2005) 3
  [arXiv:hep-th/0406092].

\bibitem{BDFS}
F.~Buccella, J.-P.~Derendinger, S.~Ferrara and C.~A.~Savoy,
  Phys.\ Lett.\ B {\bf 115} (1982) 375; \\
 C.~Procesi and G.~W.~Schwarz,
  Phys.\ Lett.\ B {\bf 161}, 117 (1985).

\bibitem{noscale}
  E.~Cremmer, S.~Ferrara, C.~Kounnas and D.~V.~Nanopoulos,
  Phys.\ Lett.\ B {\bf 133} (1983) 61; \\
  J.~R.~Ellis, A.~B.~Lahanas, D.~V.~Nanopoulos and K.~Tamvakis,
  Phys.\ Lett.\ B {\bf 134} (1984) 429;\\
  J.~R.~Ellis, C.~Kounnas and D.~V.~Nanopoulos,
  Nucl.\ Phys.\ B {\bf 247} (1984) 373;\\
  A.~B.~Lahanas and D.~V.~Nanopoulos,
  Phys.\ Rept.\  {\bf 145} (1987) 1.




\end{thebibliography}
\end{document}